\documentclass{moriond}

\usepackage{amsfonts}
\usepackage{amsmath, amssymb}

\newcommand{\GeV}{{\ensuremath\rm GeV}}
\newcommand{\TeV}{{\ensuremath\rm TeV}}
\newcommand{\pb}{{\ensuremath\rm pb}}
\newcommand{\fb}{{\ensuremath\rm fb}}
\newcommand{\lb}{\left(}
\newcommand{\rb}{\right)}

\def\be{\begin{equation}}
\def\ee{\end{equation}}
\def\bea{\begin{eqnarray}}
\def\eea{\end{eqnarray}}

\newcommand{\Photo}{}

\begin{document}
\rightline{RBI-ThPhys-2023-12, CERN-TH-2023-081}
\vspace*{4cm}
\title{TWO-REAL-SINGLET MODEL BENCHMARK PLANES\\ \vspace{3mm} - A MORIOND UPDATE -}

\author{T. ROBENS}

\address{Ruder Boskovic Institute, Bijenicka cesta 54, 10000 Zagreb, Croatia\\and\\ Theoretical Physics Department, CERN, 1211 Geneva 23, Switzerland}

\maketitle\abstracts{
I present an update on the the Benchmark Planes in the Two-Real-Singlet Model (TRSM), a model that enhances the Standard Model (SM) scalar sector by two real singlets, where an additional $\mathbb{Z}_2 \otimes \mathbb{Z}'_2$ symmetry is imposed. I discuss the case where all fields acquire a vacuum expectation value, such that the model contains in total 3 CP-even neutral scalars that can interact with each other. I remind the readers of the previously proposed benchmark planes, current constraints, and possible signatures at current and future colliders. This is an update for Moriond 2023 of results presented in \cite{Robens:2022nnw}.}

\section{Introduction and Model}
The model discussed here has already been widely discussed in the literature \cite{Robens:2019kga,Papaefstathiou:2020lyp,Robens:2022nnw}, to which we refer the reader for more details. We here just briefly recapitulate the main characteristics of the model.

The Two-Real-Singlet Model (TRSM) is a new physics model that enhances the Standard Model (SM) electroweak sector by two additional fields that are singlets under the SM gauge group. The fields obey an additional $\mathbb{Z}_2\,\otimes\,\mathbb{Z}_2'$ symmetry.  All scalar fields acquire a vacuum expectation value and therefore mix with each other. One of the resulting three CP even neutral scalars has to comply  with the measurements of the Higgs boson by the LHC experiments \cite{ATLAS:2022vkf,CMS:2022dwd}.

In \cite{Robens:2019kga}, several processes were defined that were by that time not investigated at the LHC, which can be classified as
either asymmetric (AS) production and decay, in the form of
$p\,p\,\rightarrow\,h_3\,\rightarrow\,h_1\,h_2$,
or
symmetric (S) decays in the form of
$p\,p\,\rightarrow\,h_i\,\rightarrow\,h_j\,h_j$,
where in our study none of the scalars corresponds to the 125 \GeV~ resonance.  In the following, we use the convention that $M_1\,\leq\,M_2\,\leq\,M_3$ for the masses of the scalars $h_{1,2,3}$ in the mass eigenstates. We here consider the following benchmark planes (BPs), where cross sections refer to production at a 13 \TeV~ $pp$ collider:

\begin{itemize}
  \item {\bf   AS} {\bf   BP1: $h_3 \to h_1 h_2$ ($h_3 = h_{125}$) }: SM-like decays for both scalars: $\sim\,3\,\pb$; $h_1^3$ final states: $\sim 3\, \pb$
 \item{} {\bf   AS} {\bf   BP2: $h_3 \to h_1 h_2$ ($h_2 = h_{125}$)}: {SM-like decays for both scalars: $\sim\,0.6\,\pb$
}
\item {\bf   AS} {\bf   BP3: $h_3 \to h_1 h_2$ ($h_1 = h_{125}$)}: { (a) SM-like decays for both scalars $\sim\,0.3\,\pb$; {  (b) $h_1^3$ final states: $\sim\,0.14\,\pb$}}
  \item {\bf   S} {}{\bf   BP4: $h_2 \to h_1 h_1$ ($h_3 = h_{125}$)}: {up to 60 \pb}
  \item {} {\bf   S} {\bf   BP5: $h_3 \to h_1 h_1$ ($h_2 = h_{125}$)}: { up to $2.5\,\pb$}
 \item {} {\bf   S} {\bf   BP6: $h_3 \to h_2 h_2$ ($h_1 = h_{125}$)}:
{SM-like decays: up to 0.5 \pb; {$h_1^4$ final states: around 14 \fb}}
\end{itemize}

\section{Updated benchmark planes}
The main result presented here is an update of the benchmark planes from \cite{Robens:2019kga,Robens:2022nnw} to include more recent search constraints. For this, we have developed an interface to HiggsTools \cite{Bahl:2022igd}. The following searches have led to additionally excluded regions in the benchmark planes:
\begin{itemize}
\item{}a full Run II result for $H\,\rightarrow\,Z\,Z\,\rightarrow\,\ell\ell\ell\ell$ from the ATLAS collaboration \cite{ATLAS:2020tlo}
\item{}a full Run II result for $H\,\rightarrow\,h_{125}\,h'\,\rightarrow\,\tau^+\tau^-\,b\,\bar{b}$ from the CMS collaboration \cite{CMS:2021yci}.
\end{itemize}
Especially the latter, in principle a prime example for a search covering one of the asymmetric benchmark planes in our model, has only been interpreted in an NMSSM scenario in the experimental publication, although some regions of our benchmark planes prove to be sensitive. I therefore stronly encourage the experimental collaborations to also interpret their bounds within the TRSM in future searches. Furthermore, for these proceedings I changed exclusion criteria within HiggsTools from the expected sensitivity, which is the default within this tool, to the observed sensitivity. Therefore, some additional regions in the parameter space might now be ruled out by searches already considered in the references given above.

In the following, $h_3\,\rightarrow\,h_1\,h_2$ labels regions in the parameter space excluded by \cite{CMS:2021yci}, while $h_3\,\rightarrow\,h_1\,h_1$ correspond to results from \cite{CMS:2018qmt,CMS:2018ipl,ATLAS:2019qdc} and $h_3\,\rightarrow\,Z\,Z$ from \cite{ATLAS:2020tlo}, respectively.

We see that in particular the $H\,\rightarrow\,Z\,Z$ search rules out large regions in the models parameter space; for BP2, basically masses $M_3\,\in\,\left[355;380\right]\,\GeV$ are now excluded. Similarly, a large region in BP3 is sensitive to this search, in particular in the region where $M_3\,-\lb M_1+M_2 \rb$ is small such that the $ZZ$ branching ratio gets enhanced. In addition, for all BPs presented here the asymmetric search $H\,\rightarrow\,h_{125}\,h'\,\rightarrow\,\tau^+\tau^-\,b\,\bar{b}$ \cite{CMS:2021yci} constrains some regions in the parameter space, typically in regions where $h_1$ and $h_2$ are close in masses. In BP6, an additional exclusion stems from the ATLAS early Run II search for $H\,\rightarrow\,H'\,H'\,\rightarrow\,W^+\,W^-\,W^+\,W^-$ \cite{ATLAS:2018ili}.

\begin{center}
\begin{figure}[tbh]
\begin{center}
\includegraphics[width=0.45\textwidth]{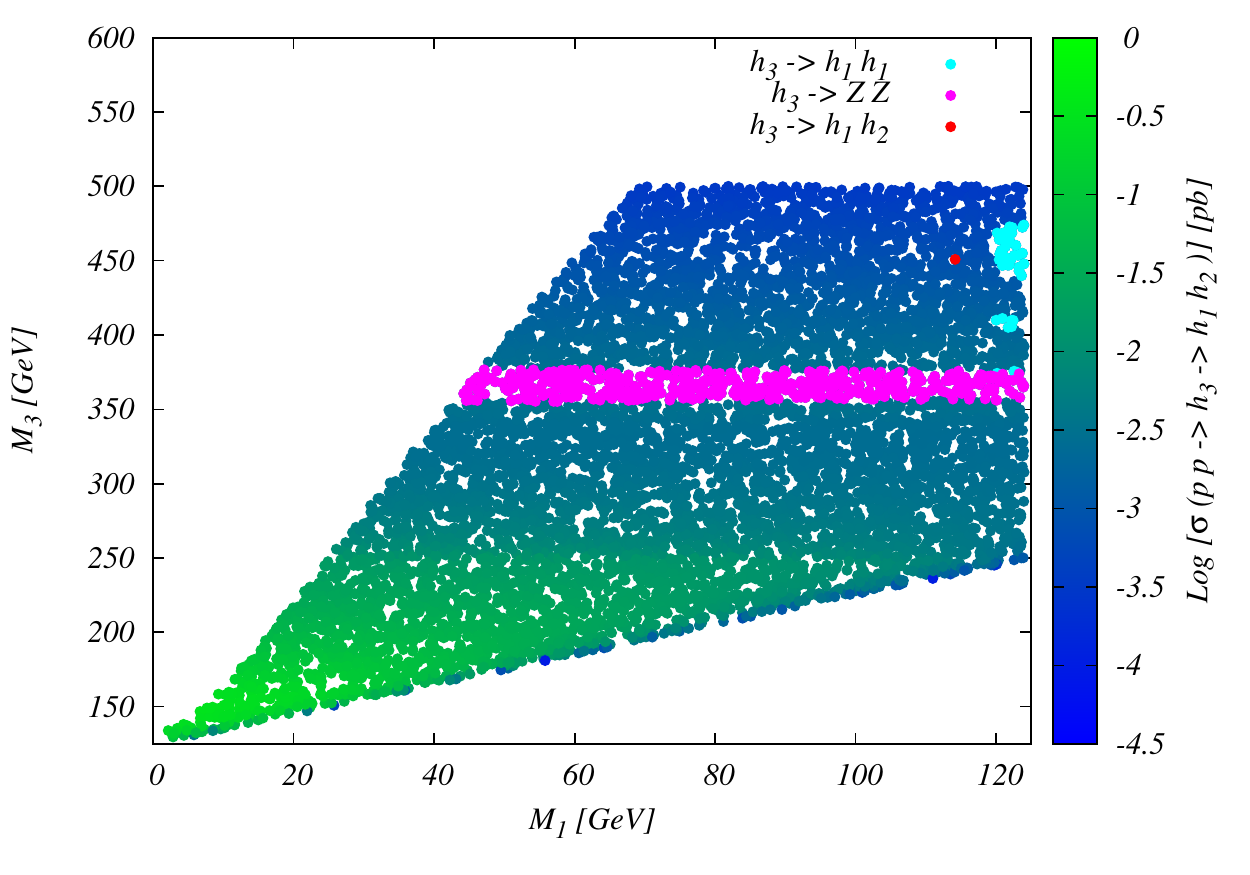}
\includegraphics[width=0.45\textwidth]{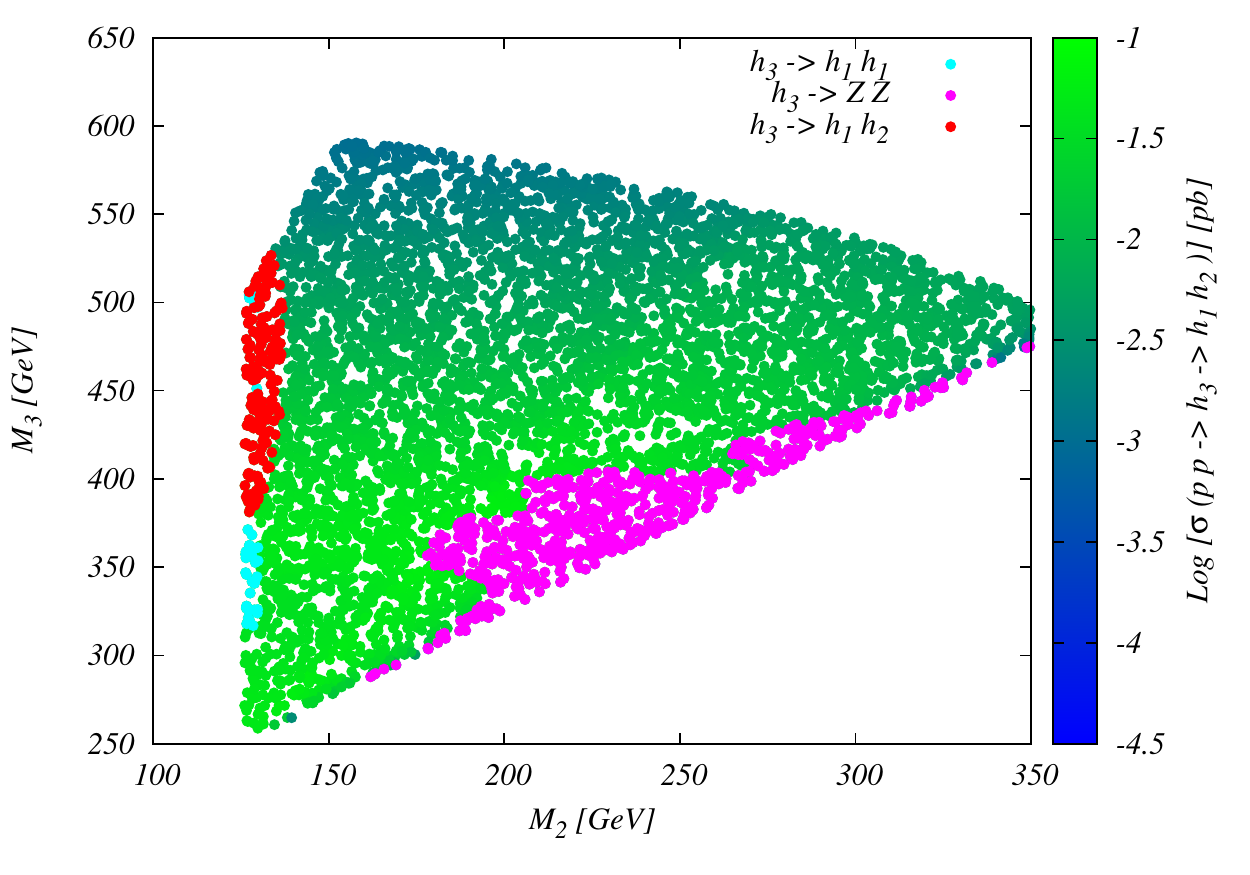}
\caption{Parameter regions in BP2 {\sl(left)} and BP3 {\sl (right)} with updated constraints; see text for details.}
\end{center}
\end{figure}
\end{center}
\begin{center}
\begin{figure}[tbh]
\begin{center}
\includegraphics[width=0.45\textwidth]{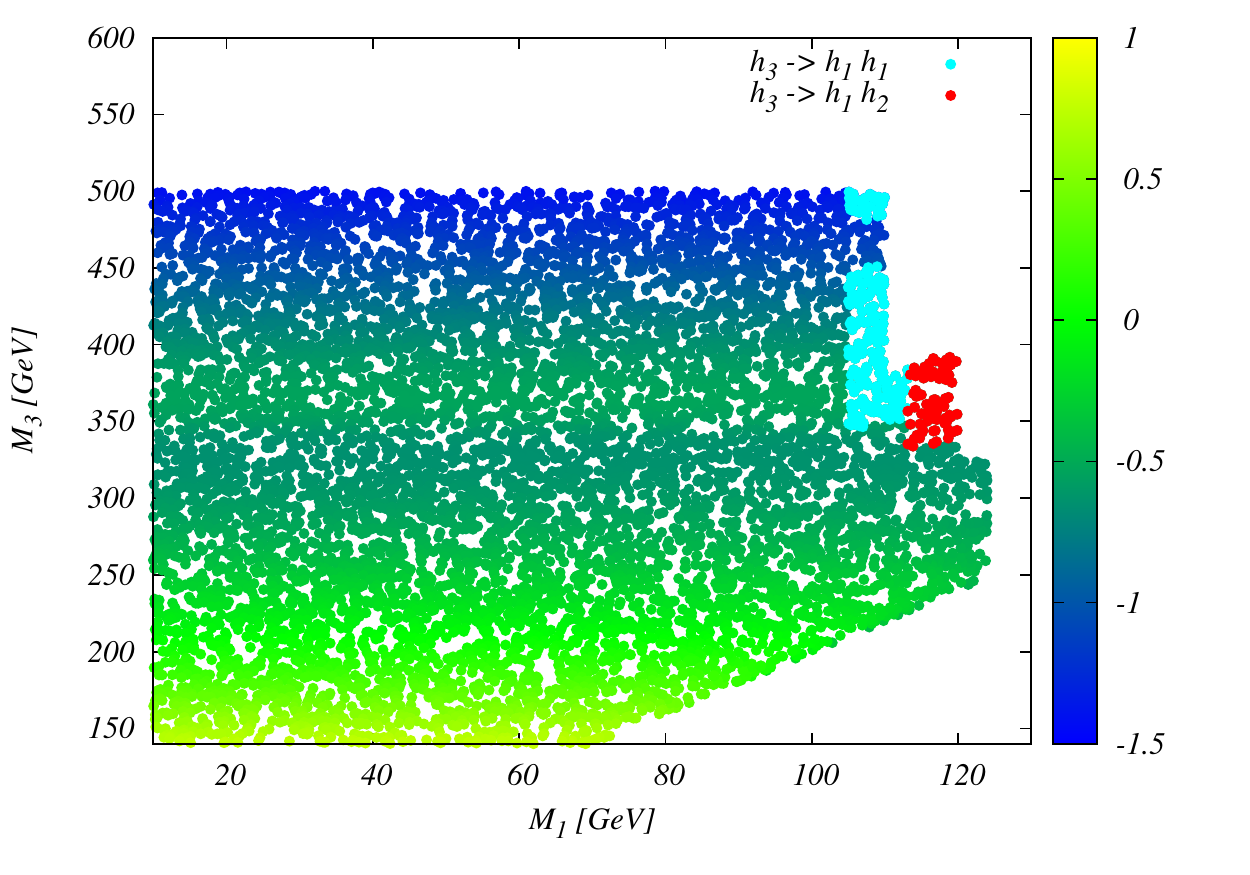}
\includegraphics[width=0.45\textwidth]{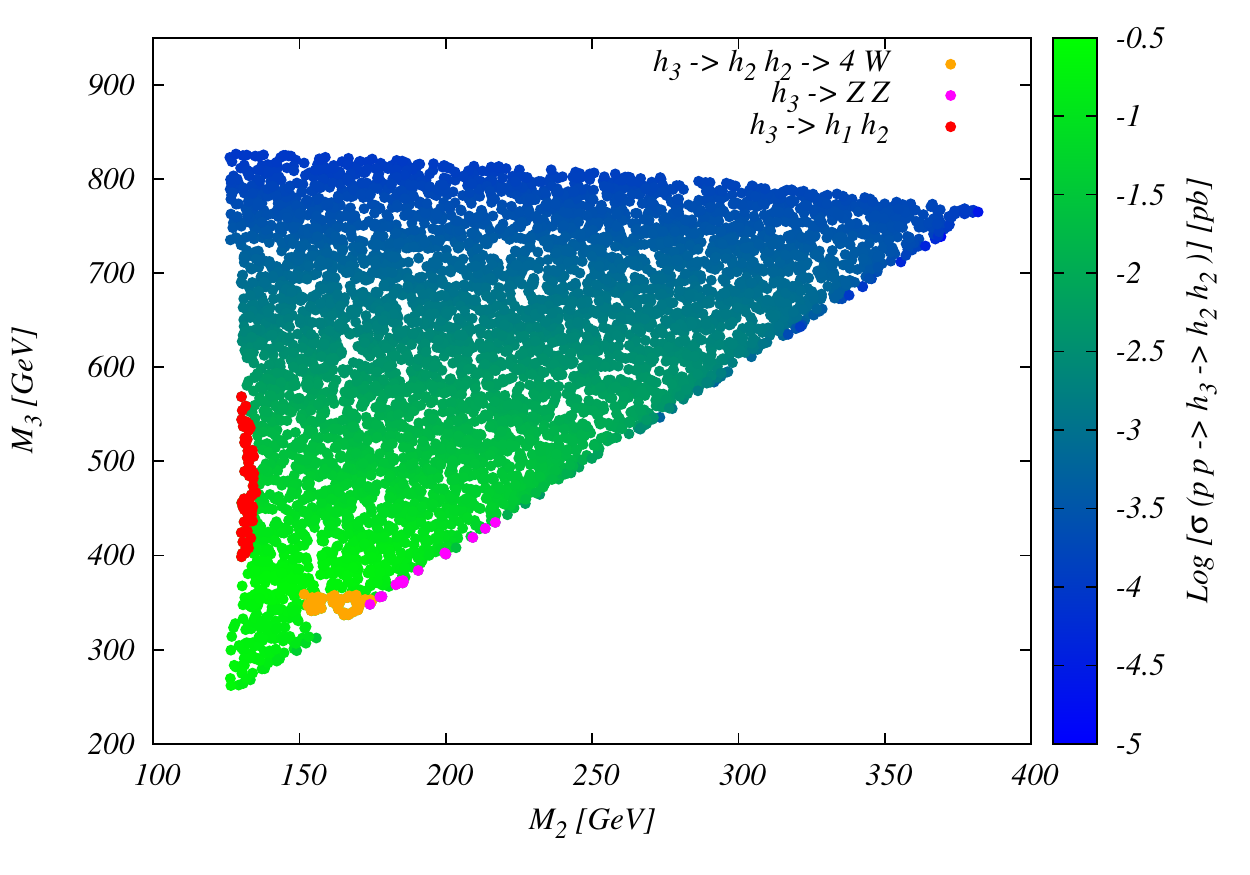}
\caption{Parameter regions in BP5 {\sl(left)} and BP6 {\sl (right)} with updated constraints; see text for details.}
\end{center}
\end{figure}
\end{center}
\section{LHC interpretations}

\begin{center}
\begin{figure}[tbh]
\begin{center}
\includegraphics[width=0.9\textwidth]{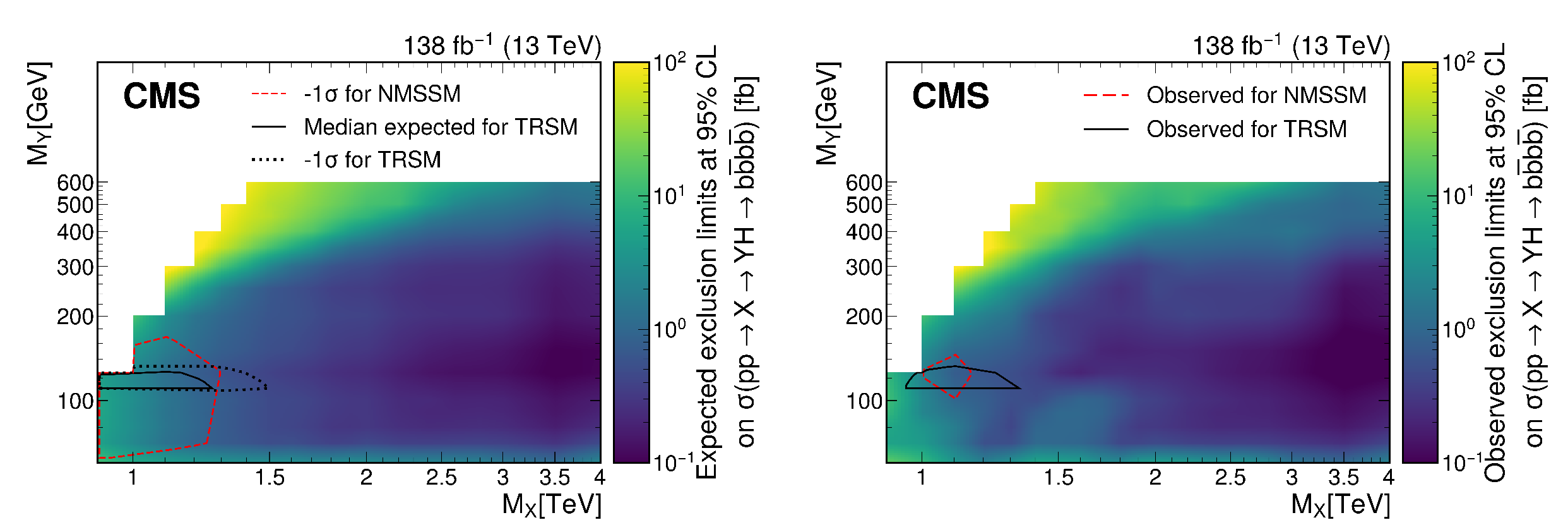}
\caption{\label{fig:cmsres} Expected {\sl (left)} and observed {\sl (right)} $95\%$ confidence limits for the $p\,p\,\rightarrow\,h_3\,\rightarrow\,h_2\,h_1$ search, with subsequent decays into $b\bar{b}b\bar{b}$. For both models, maximal mass regions up to $M_3\,\sim\,\,1.4\,\TeV,\;M_2\,\sim\,\,140\,\GeV$ can be excluded.}
\end{center}
\end{figure}
\end{center}

\begin{center}
\begin{figure}[tbh]
\begin{center}
\includegraphics[width=0.45\textwidth]{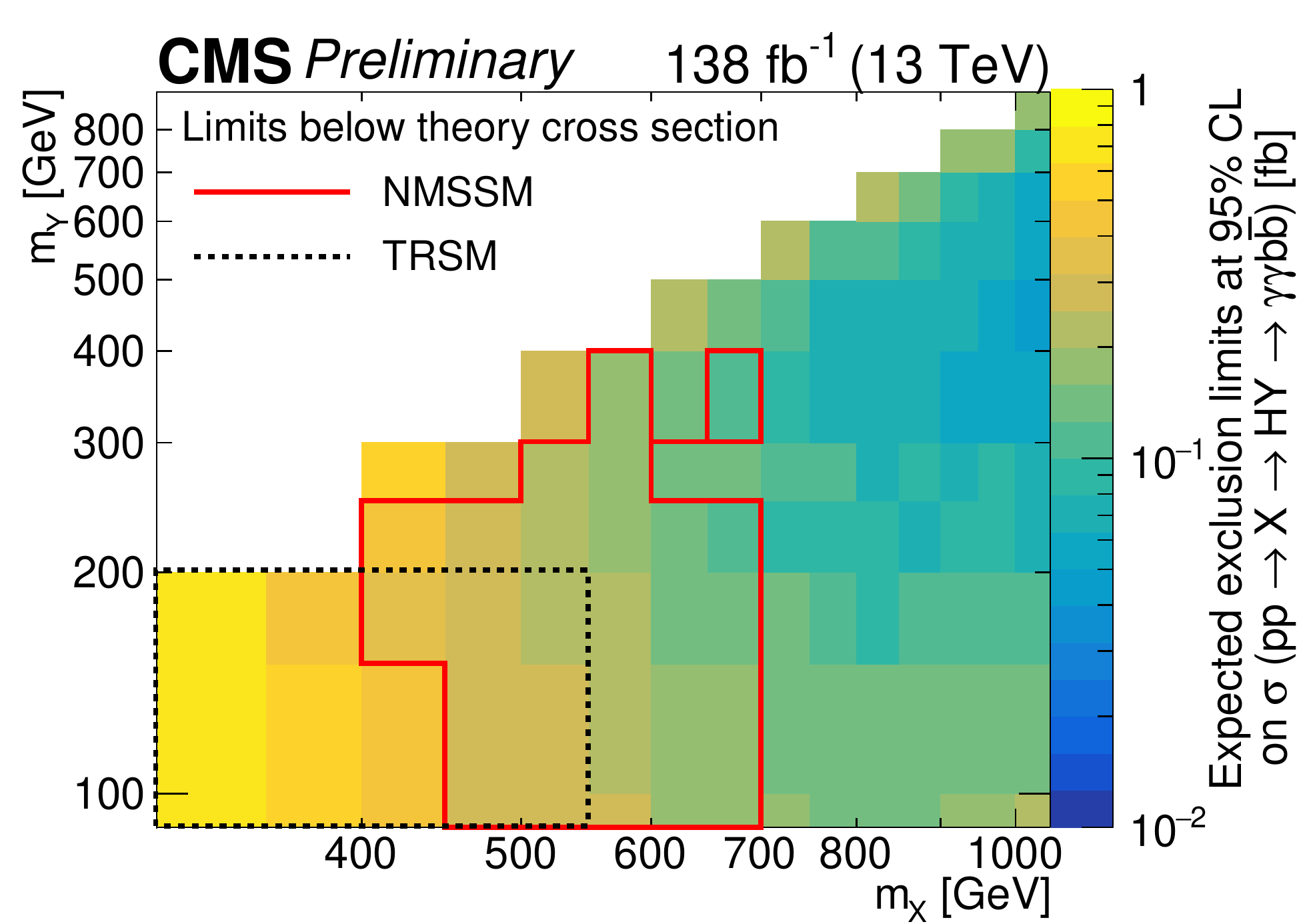}
\includegraphics[width=0.45\textwidth]{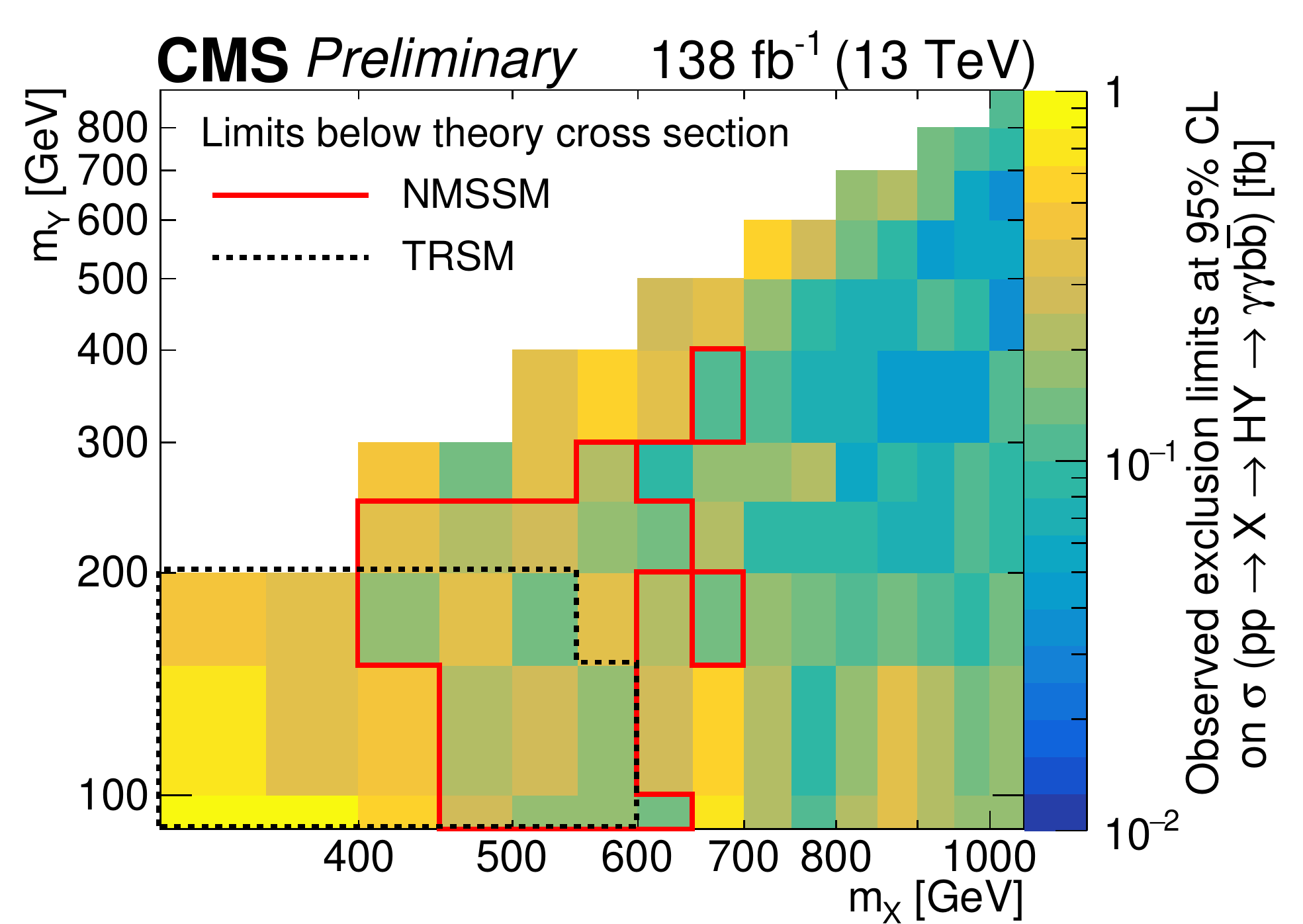}
\caption{\label{fig:cmsbbgaga} Expected {\sl (left)} and observed {\sl (right)} $95\%$ confidence limits for the $p\,p\,\rightarrow\,h_3\,\rightarrow\,h_2\,h_1$ search, with subsequent decays into $b\bar{b}\gamma\gamma$. Depending on the model, maximal mass regions up to $m_3\,\sim\,\,800\,\GeV,\;m_2\,\sim\,\,400\,\GeV$ can be excluded.}
\end{center}
\end{figure}
\end{center}

Two experimental searches have by now made use of the predictions obtained within the TRSM to interpret regions in parameter space that are excluded: a CMS search for asymmetric production and subsequent decay into $b\bar{b}b\bar{b}$  final states \cite{CMS:2022suh}, as well as $b\bar{b}\gamma\gamma$ in \cite{CMS-PAS-HIG-21-011}. Maximal rates for these within the TRSM are documented in \cite{reptr,trsmbbgaga}. Figures \ref{fig:cmsres} (taken from \cite{CMS:2022suh}) and \ref{fig:cmsbbgaga} (taken from \cite{CMS-PAS-HIG-21-011}) show the expected and observed limits in these searches for the TRSM and NMSSM \cite{Ellwanger:2022jtd}.

\section{Conclusions}
The TRSM is a new physics model that, with a small number of additional new physics parameters, allows for novel final states, in particular asymmetric scalar production and decays. I here presented updated benchmark planes for this model, where novel experimental constraints with respect to \cite{Robens:2022nnw} have been included. I strongly encourage the experimental collaborations to reinterpret their searches within this model. UFO file and maximal cross section values are available upon request.
\section*{Acknowledgments}
I thank the organizers for repeated financial support and for choosing a fantastic location inspiring group spirit.
\section*{References}

\end{document}